# Analytical and FDTD design of hybrid Fabry-Perot plano-concave microcavity for hexagonal boron nitride

Felipe Ortiz-Huerta,[1,*] Karina Garay-Palmett.[1]

[1] *Departamento de Óptica, Centro de Investigación Científica y de Educación Superior de Ensenada, Apartado Postal 2732, BC 22860 Ensenada, México.*

The emission of light of an efficient single-photon emitter (SPE) should have a high-count rate into a well-defined spatio-temporal mode along with an accessible numerical aperture (NA) to increase the light extraction efficiency that is required for effective coupling into optical waveguides. Based on a previously developed experimental approach to fabricate hybrid Fabry-Perot microcavities (Further information is available [F. Ortiz-Huerta et al, Opt.Express **26**, 33245 (2018)]), we managed to find analytical and finite-difference time-domain (FDTD) values for the, experimentally achievable, geometrical parameters of a hybrid plano-concave microcavity that enhances the spontaneous emission (i.e. Purcell enhancement) of color centers in two-dimensional (2D) hexagonal boron nitride (hBN) while simultaneously limiting the NA of the emitter. Paraxial approximation and a transfer matrix model are used to find the spotsize of the fundamental Gaussian mode and the resonant modes of our microcavity, respectively. A Purcell enhancement of 6 is found for a SPE (i.e. in-plane dipole) hosted by a 2D hBN layer inside the hybrid plano-concave microcavity.

## I. INTRODUCTION

Pure and indistinguishable SPEs are key components needed for their application in upcoming quantum technologies [1] (e.g. quantum computation [2], quantum networks [3]).Color centers in 2D hBN and diamonds are among the most promising candidates for solid-state single-photon emission at room temperature [4,5]. Nonetheless, in contrast with bulk diamond, the 2D nature of hBN, hosting color centers (i.e. in-plane dipoles), overcomes the necessity for geometrical approaches [6] (i.e. solid immersion lenses [7]) to reduce the angle of emission of the selected SPE.

Challenges still lie ahead for hBN as an ideal SPE [4] and, in order to overcome them, photonic structures such as open-access Fabry-Perot microcavities [8] and photonic crystals [9] have been built around color centers in hBN to increase its spontaneous emission by means of Purcell effect. An alternative and low-cost approach to build photonic structures uses polymers to embed different types of SPEs (e.g. quantum dots [10], molecules [11]) by a process known as two-photon polymerization (2PP) [12] where a photopolymer resist is illuminated with a focused laser at 780nm and absorbs two photons simultaneously, which triggers a corresponding chemical reaction that solidifies the material to build the desired shape.

A natural extension to the development of polymer photonic structures consists of the fabrication of hybrid (i.e. metal -dielectric) resonant structures [13] with the potential to enhance the light-matter interactions of such SPEs. This work will focus on finding an optimal design for a hybrid plano-concave microcavity, containing a monolayer of hBN hosting a SPE (Fig. 1), by using analytical methods and FDTD simulations.

\_\_\_\_\_\_\_\_\_

*fortiz.huerta@gmail.com

Fabrication design steps are first shown for our microcavity (Section II A), afterwards we found the range of geometrical parameters necessary for our stable resonator (Section III A), followed by a transfer matrix model used to find the resonant modes of the microcavity (Section III B), which are then corroborated by FDTD simulations (Section IV A).

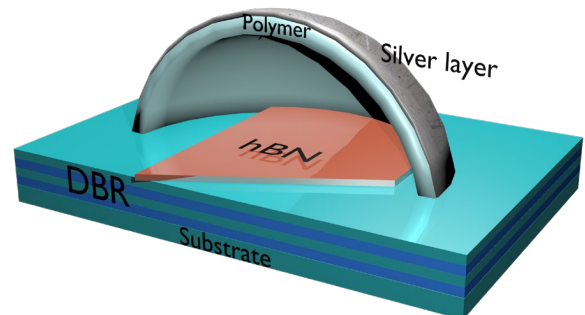

FIG. 1. Conceptual design shows cross-section of hybrid plano-concave microcavity with a 2D hBN layer inside on top of a distributed Bragg reflector (DBR).

## II. FABRICATION DESIGN

### A. Hybrid plano-concave microcavity

By using a quarter-wavelength DBR (Section 3B) with a multilayer 2D material on top [Fig. 2(a)], we designed our system (2D material+ DBR stack) to have a maximum reflectivity at the center wavelength of 637nm (section 3B). The selected wavelength of our system falls within the typical emission rates of the zero-phonon line (ZPL) of SPEs in hBN (500-800nm). A quarter-wavelength thickness is conveniently chosen for the hBN (Section III B) where its value falls between experimentally achievable thicknesses of multilayer 2D materials [8].

A 3D concave shape polymer then could be fabricated on top of the 2D material [Fig. 2(b)] by a direct laser writing system (e.g. Photonic Professional, Nanoscribe GmbH) by use of a 2PP process.

Afterwards an 80nm silver layer could be added, by thermal evaporative deposition, on top of the concave shape polymer to ensure a high reflectivity inside our microcavity. When designing the concave shape polymer a small rectangular aperture at its edge must be taken into account in the fabrication step [Fig.2(b),(c)] to prevent the accumulation of the photopolymer resist inside the solidified concave polymer after the 2PP process is finished.

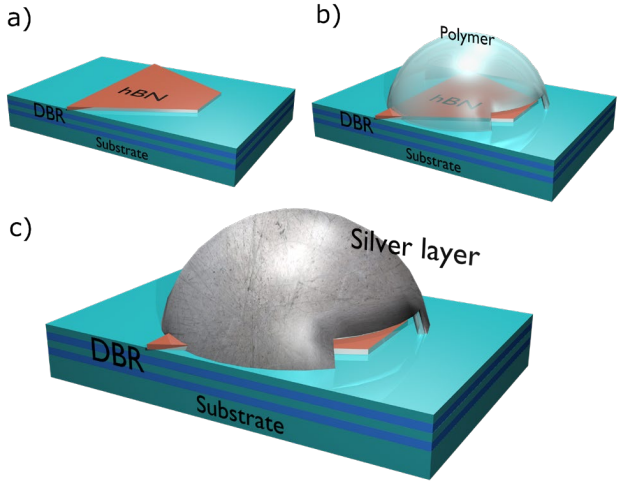

FIG. 2. Fabrication steps of hybrid microcavity. (a) hBN layer positioned on top of DBR. (b) Concave polymer shape is fabricated by direct laser writing process. (c) A silver layer is added on top of polymer.

## III. ANALYTICAL DESIGN

### A. Geometrical Parameters of Plano-Concave Microcavity

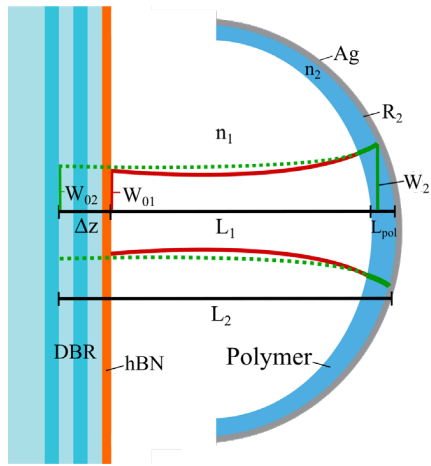

FIG. 3: Cross-section of hybrid plano-concave microcavity shows the geometrical parameters and the two Gaussian modes inside.

When a polymer layer is added inside a bare microcavity, as in our case, two fundamental Gaussian beams are formed inside the air gap and polymer layer respectively (Fig. 3) [14].

The spotsize $W_{02}$ (Fig. 3) of the fundamental Gaussian mode ($TEM_{00}$) inside the cavity has to be as small as possible, since this means a small modal volume and consequently, a high Purcell factor [15].

By setting an arbitrary range of values for the length of the second Gaussian beam $L_2$ and radius of curvature $R_2$ of our plano-concave microcavity, Fig. 4 shows the spotsizes $W_{02}$ and $W_2$ corresponding to different pair of values ($R_2, L_2$) for a hybrid plano-concave cavity. The spotsizes $W_{02}$ and $W_2$ are calculated by [16]:

$$W_{02}^2 = \frac{L_2 \lambda_0}{\pi n_2} \sqrt{\frac{g}{1-g}} \quad (1)$$

and

$$W_2^2 = \frac{L_2 \lambda_0}{\pi n_2} \sqrt{\frac{1}{g(1-g)}}, \quad (2)$$

respectively, where $g = 1 - L_2/R_2$ is the stability range for our plano-concave cavity and $\lambda_0 = 637nm$ is the wavelength of the fundamental Gaussian mode, $n_2 = 1.52$ is the refractive index of the polymer layer. The length of the second Gaussian beam is defined as $L_2 = L_1 + L_{pol} + \Delta z$, where $L_1$ is the length of the Gaussian beam in air, $L_{pol}$ is the polymer thickness and $\Delta z$ is calculated by the ABCD law [14]:

$$q_2 = \frac{Aq_1 + B}{Cq_1 + D}, \quad (3)$$

where the complex numbers $q_{1,2} = z_{1,2} + jz_{R,1,2}$ are known as the q-parameters for the Gaussian beams, where $z_2 = L_2 - L_p$, $z_1 = L_1$ and $z_{R,1,2}$ is the Rayleigh length for each beam. For a Gaussian beam passing through a plane dielectric interface, we have $A = B = C = 0$, and $D = n_2/n_1$, where $n_1 = 1$ is the refractive index of the air gap, therefore, by substituting in eq. (3), $q_2 = (n_2/n_1)q_1$. This leads to $z_2 = (n_2/n_1)z_1$ and $W_{01} = W_{02}$. Finally, by defining $\Delta z = z_2 - z_1$ we get:

$$\Delta z = \left(\frac{n_2}{n_1} - 1\right) L_1. \quad (4)$$

As a threshold for $R_2$ we set $R_2 \geq L_2$ in accordance with the stability range where $0 \leq g \leq 1$. Although work has been done to include the lensing effect of a curved "$n_1/n_2$" interface (see supplementary material of [17]), the planar surface ($R_1 = \infty$) approximation values (Table 1) fall within the desired range with our FDTD simulations (Section IV A).

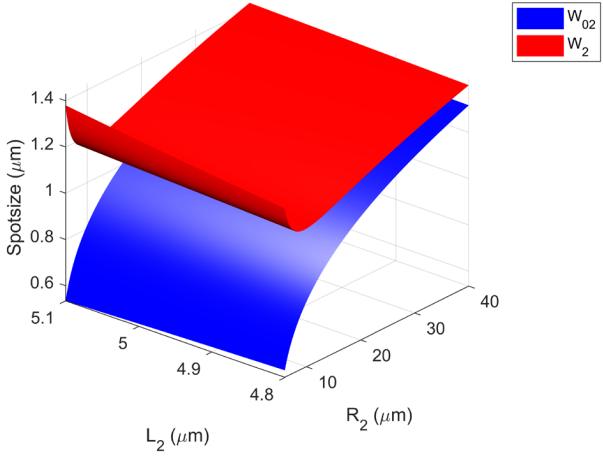

FIG. 4. Spotsizes $W_0$ and $W_{02}$ for different values of $R_2$ and $L_2$.

We take a transversal cut through a fixed value of $L_2$ (Fig. 5) and observe the dependence of $W_{02}$ and $W_2$ to the radius of curvature ($R_2$) of a plano-concave cavity. To achieve a high Purcell factor, and a small NA, $R_2$ must be as small as possible (small $W_{02}$), while maintaining the lower boundary condition ($R_2 \geq L_2$), therefore the optimal values of $R_2$, for any arbitrary $L_2$, will resided near the vicinity of the minima of the $W_2$ function(Fig. 5), setting the boundary values for $R_2$, for any given $L_2$, at $R_2 \approx 2L_2$.

Selecting the $R_2$ parameter closer to the divergence of the $W_2$ function ($R_2 = L_2$) could result in unstable resonators that will not hold a stable Gaussian mode inside. Theoretical work has been done with $R_2 \approx L_2$, [18] where a non-paraxial analysis is performed, although diffraction losses have to be considered for an accurate description of the experimental limits of stability [19]. In the unstable regime ($R_2 < L_2$) extensive work has also been developed [20, 21].

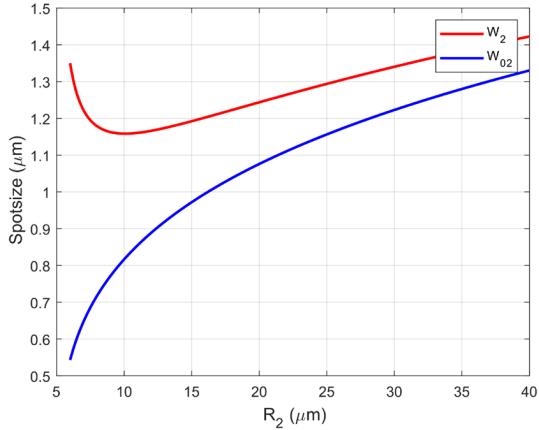

FIG. 5. Transverse cut of Fig. 4 through length $L_2 = 5.03 \mu m$ to show dependence of $R_2$ with spotsizes. As the values of $R_2$ diminishes, while maintaining a constant $L_2$, the functions for $W_{02}$(blue) and $W_2$(red) start to diverge, arriving at the limit of the paraxial approximation (stability regime).

### B. Electric Field Distribution and Resonant modes of Plano-Concave Microcavity

A $\lambda_0/4n$ thickness layer of hBN ($n = 1.72$) was positioned on top of a 15-pair layer DBR with tantalum oxide ($Ta_2O_5$) and silicon oxide ($SiO_2$) as the High- and Low-index layers, respectively, on a $(HL)^{15}$ configuration to ensure an Electric field antinode at the surface of the hBN layer, making the hBN + DBR system a $L(HL)^{15}$ dielectric stack. A transfer matrix model [22] was used to calculate the electric field distribution inside the hBN + DBR system (Fig. 6).

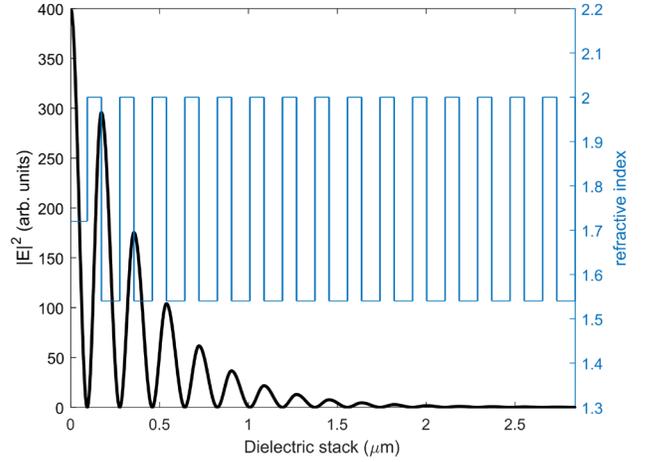

FIG. 6. Electric field distribution of a hBN + DBR system on a $L(HL)^{15}$ configuration. Maximum Electric field intensity is found at the surface of the hBN layer. Vertical lines (blue) represent the boundaries between each dielectric layer.

The full transfer matrix $S$ of our microcavity is defined as:

$$S = L_{Ag}I_1 L_{pol} I_2 L_{air} I_3 L_{hBN} I_4 L_{DBR} I_5, \quad (5)$$

where $L$ and $I$ represent the transfer and interface matrix, respectively, of the silver (Ag), polymer (pol), air, hBN and DBR layer. The transfer matrices $L_{pol}$ and $L_{air}$ are defined as [23]:

$$L_{pol} = \begin{bmatrix} exp\left(-\frac{i2\pi n_2}{\lambda_0} + iG_2\right) & 0 \\ 0 & exp\left(\frac{i2\pi n_2}{\lambda_0} - iG_2\right) \end{bmatrix}, \quad (6)$$

$$L_{air} = \begin{bmatrix} exp\left(-\frac{i2\pi n_1}{\lambda_0} + iG_1\right) & 0 \\ 0 & exp\left(\frac{i2\pi n_1}{\lambda_0} - iG_1\right) \end{bmatrix}, \quad (7)$$

where $G_{1,2} = arctan(L_{1,2}\lambda_0/n_{1,2}\pi W_{01,02})$ is the Guoy phase shift in the air $(n_1 = 1)$ and polymer layer, respectively, where $W_{01} = W_{02}$.

The transmittance of the microcavity is calculated, from the matrix elements of $S$, to find its fundamental TEM resonant modes (Fig. 7). We found the desired TEM modes at $R_2 = 8.1 \mu m$ and $L_2 = L_1 + L_{pol} + \Delta z = 5.03 \mu m$, where $L_1 = 3.09 \mu m$, $L_{pol} = 0.4 \mu m$ and $\Delta z = 1.54 \mu m$, which gives a physical cavity length of $L = L_2 - \Delta z = 3.49 \mu m$. These values fall within the stability range $R_2 \approx 2L_2$ (Section III A).

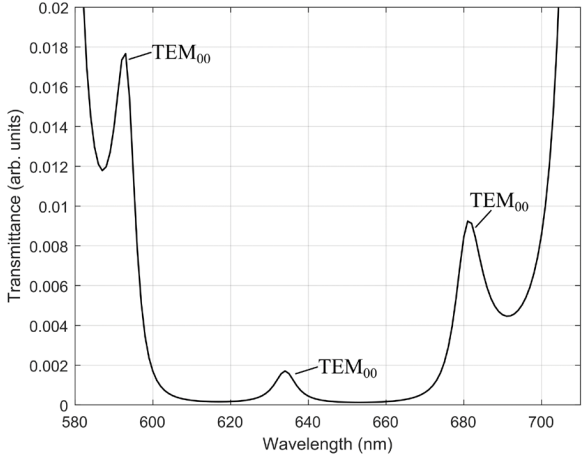

FIG. 7 Transmittance of plano-concave cavity shows the fundamental TEM modes at $595 nm$, $636 nm$ and $684 nm$.

## IV. NUMERICAL DESIGN

### A. Resonant modes of hybrid plano-concave microcavity

For the FDTD simulations, we used Ansys Lumerical FDTD software. The polymer, silver and DBR stack were treated as lossless and non-dispersive materials [13]. Identical values for the geometrical parameters previously mentioned ($R_2$, $L_2$, $L_1$), except for $R_1 = 7.7 \, \mu m$, were taken for the FDTD simulations, where an in-plane dipole emitter sits at the surface of the hBN layer to ensure a higher Purcell factor since the dipole interacts with an electric field antinode. The Purcell factor was calculated by using the classical definition [24]:

$$F_P = \frac{P_{cav}}{P_{free}}, \qquad (8)$$

where $P_{cav}$ and $P_{free}$ is the power dissipated for the dipole inside the microcavity and in free space, respectively. A Purcell factor of $F_P \approx 6$ was achieved for the TEM mode at the DBR center wavelength. A $Q$-factor of $Q = 731.4 \pm 102.7$ was also calculated in our simulations where the resonant modes of the microcavity (Fig. 8) are shown in good agreement (Table 1) with the resultant modes from the analytical model (Fig. 7).

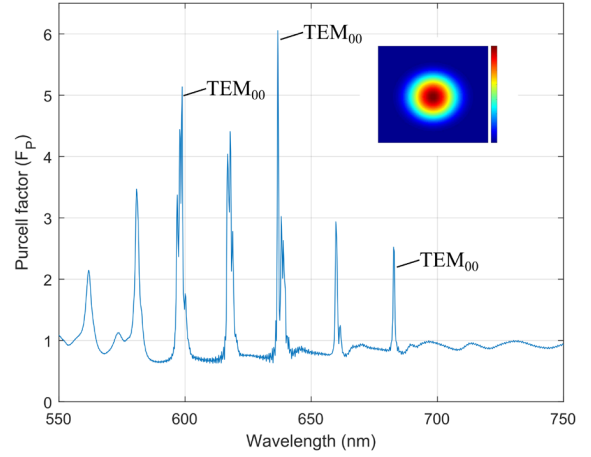

FIG. 8 Purcell factor of plano-concave microcavity. Fundamental TEM Gaussian modes are found at $595nm$, $636nm$ and $684nm$. In-set shows transverse section of fundamental Gaussian mode at 637nm.

TABLE I. Geometrical parameters and fundamental TEM mode values of designed hybrid plano-concave microcavity.

| Parameter | Analytical ($\mu m$) | FDTD ($\mu m$) |
|---|---|---|
| $R_2$ | 8.1 | 8.1 |
| Physical cavity length, L | 3.49 | 3.49 |
| $L_1$ | 3.09 | 3.09 |
| $L_2$ | 5.03 | 5.03 |
| hBN thickness | $\lambda_0/4n$ | $\lambda_0/4n$ |
| Polymer thickness | 0.4 | 0.4 |
| 1st $TEM_{00}$ | 0.595 | 0.616 |
| 2nd $TEM_{00}$ | 0.636 | 0.637 |
| 3rd $TEM_{00}$ | 0.684 | 0.684 |
| $R_1$ | $\infty$ | 7.7 |

## V. CONCLUSIONS

We have presented the fabrication design steps for a new type of hybrid plano-concave microcavity and found its fundamental resonant modes by using an expanded transfer matrix model to account for the curvature in dielectrics and, by using FDTD simulations, we were able to show the effectiveness of the analytical model and found a Purcell enhancement of 6 for a pre-selected SPE.

The geometrical parameters of our microcavity are all experimentally achievable with the two-photon absorption fabrication process [11, 13] and our modeled cavity could easily be extended to contain and enhance spontaneous emission of arbitrary solid-state SPEs [25]. Although novel approaches have been realized to diminish vibrations for open-access Fabry-Perot microcavities inside a cryostat system [26], in our design, the plano-concave microcavity is integrated directly to the substrate containing the SPE and, therefore, there are no moving parts that could potentially diminish the Purcell factor of a pre-selected SPE due to vibrations in cavity length [27], although detuning of the selected mode, due to thermally-induced contraction of the polymer by cooling [10], must be taken into account if the desired SPE and the cavity are to be analyzed inside a cryostat system.

The methodology of design of the hybrid Fabry-Perot microcavity is also suited for quantum cryptography applications, provided the emitter's wavelength is within the telecom range [6], and potential chemical sensing applications [28] , since our microcavity is also an open-access cavity.

## ACKNOWLEDGMENTS

FO acknowledges financial support from Consejo Nacional de Ciencia y Tecnología (CONACYT) postdoctoral fellowship program. KG from CONACYT Grants (Laboratorios Nacionales 315838/2021).